# Where is the main source of Jupiter family comets situated?

A.M. Kazantsev

Astronomical Observatory of Kyiv Taras Shevchenko National University, Ukraine
ankaz@observ.univ.kiev.ua

An attempt to determine spatial location of the main source of short-period comet nuclei was made. There were carried out numerical calculations for orbit evolution of Jupiter's family comets, comets with middle-period orbits and bodies of Centaur's group. On the basis of the calculations it was shown, that orbital evolution of the solar system small bodies is mainly going in the direction of the semi-major axes increase. It belongs to the bodies which can undergo approaches the planets, and orbital evolution of which is mainly going due to the gravitational forces. Such result is confirmed by qualitative analysis of changes of small body semi-major axes under approaches the planets. The conclusion was drawn that the main source of nuclei of Jupiter's family comets is apparently situated at distances from the Sun not more than 6 AU.

## 1. Introduction. The goal setting

The issue of short-period comets origin remains unsolved yet. To date comets with more long periods (middle-period comets, MdPCs) and Centaurs are considered as the most possible sources of Jupiter's family comets (JFCs). Besides the asteroid population is now supposed as a possible origin of some part of JFCs. The existence of the Main Belt Comets is a forcible argument for such opinion. In addition, there are indications that spectral characteristics of D-type asteroids (prevailing among Hilda group) are similar to the cometary ones (di Sisto Romina P. et. al. 2005). The similar characteristics should have the Jovian Trojans as well because the majority of them belong to the *D*-taxonomic class (Roig, F. et. al. 2008). Thus supposed sources of JFCs are situated at highly different distances from the Sun.

All possible sources of JFCs may be divided on two groups: the outer and the interior. It is suitable to set the boundary between the groups at distance from the Sun of 6 AU. Centaurs and MdPCs belong to the first group. Asteroids of different populations (at the outer zone of MBA, the Hilda-group, the Jovian Trojans) belong to the second one. The task lay in answer the question: what group of JFCs sources (the outer or the interior) is the main source?

## 2. Approach to the task solution

We want to solve this problem by numerical integration of JFC's orbits and orbits of bodies from the external sources.
In order to solve the task it should take the following steps:
1. To select JFC orbits with aphelion distances $Q < 6$ AU.
2. To select all observable MdPCs and Centaurs.
3. To estimate numbering of all existing (not only observable) MdPCs and Centaurs with sizes correspond to the sizes of JFC nuclei.
4. To calculate orbit evolution of JFC, MdPCs and Centaurs for long time periods. During the evolution some part of JFCs will leave the family. Some part of MdPCs and Centaurs will move on orbits typical for JFCs.
5. It is necessary to estimate and compare the velocities of outflow and replenishment of JFCs. On the base of calculation results to make a conclusion on the possibility to sustain an existing number of JFCs due to replenishment from the external sources.

## 3. Selection for JFCs, MdPCs and Centaurs

There were selected JFCs with aphelion distances $Q < 6$ AU. Such restriction is caused by above-mentioned division principle of the Solar system into two zones (the outer and the interior). At the end of 2011 the total comet quantity was equal 210. The set consist of all comets which were observed at the latest time not earlier than two orbital periods ago.

There were selected all observable MdPCs with aphelion distances $Q > 6.1$ AU and with orbital periods $P < 200$ years. At the end of 2011 the total MdPCs quantity was equal 211. Because the MdPCs can move much farther from the Earth and from the Sun than JFCs, not of all existing MdPCs may be opened by that time. To estimation numbering of undiscovered MdPCs it may consider distribution of the MdPCs upon perihelion distances q (tabl. 1). One can see from the table that quantity of the MdPC with $q < 5$ AU is 195, i.e., 92% of the sample. All JFCs are moving on orbits with perihelion distances $q < 5$ AU. The comet activity at such distances from the Sun is easily detected by modern observational equipment. Therefore undiscovered MdPCs may move on orbits with q > 5 AU only. However, at such heliocentric distances a few comets can display the comet activity, and majority of comet nuclei are registering as the Centaurs. Consequently number of all existing MdPCs evidently differ little from the number of opened ones.

Table 1. Distribution MdPC on the perihelion distances

| $q$, AU | $N_q$ |
|---|---|
| 0.0 – 1.0 | 18 |
| 1.0 – 2.0 | 70 |
| 2.0 – 3.0 | 65 |
| 3.0 – 4.0 | 19 |
| 4.0 – 5.0 | 23 |
| 5.0 – 6.0 | 10 |
| 6.0 – 10.0 | 5 |
| 10.0 – 12.0 | 1 |

Nevertheless 211 artificial MdPCs were added to the 211 real ones. All orbital elements of any artificial comet corresponded to all orbital elements one of the real comets, with the exception of the mean anomaly which differed by 180° from the corresponding real value. In according to above explanation, such increase of MdPCs quantity is lot more than it is necessary for compensation of the observational selection.

There are no very distinct orbit parameters for Centaurs. For selection of such bodies the next orbital parameters were accepted: semi-major axes of 6 – 25 AU, eccentricities of 0.0 - 1.0, inclinations of 0° – 180°, aphelion distances of 6.5 - 30 AU. There are 70 Centaurs with above orbital elements in the MPC catalogue at the end of 2011.

Estimation of all existing Centaurs quantity with certain sizes is more complicated matter. According to Snodgrass C. et. al. (2011) one can accept the minimal size (diameter) of JFCs is equal 1 km. Asteroid sizes are calculated on the absolute magnitude $H$ and geometric albedo $p_v$ of the body. For the IRAS catalogue (Tedesco et. el. 2002) the next expression was used

$$2 \lg D(\text{km}) = 6.247 - 0.4H - \lg p_v \qquad (1)$$

Since the Centaur's sizes and their albedos are not pointed in the catalogue, the sizes can be estimated approximately by on mean albedo for the group. The mean albedo for Centaurs is equal about 0.08 (Stansberry J. et. al. 2008; Fornasier S. et. al. 2012). Thus expression (1) may be rewritten in the next view

$$D(\text{km}) = 10^{0.2(18.3 - H)} \qquad (2)$$

One can see from the (2) that absolute magnitude of a Centaur with $D = 1$ km is about $18.3^m$.

For number estimations of all existing such bodies one can use size-distribution formulae

$$dN(D) = kD^{-b}dD \qquad (3)$$

where $dN(D)$ – asteroid quantity in a narrow size site ($dD$), $k$ and $b$ – certain constant parameters. The same formulae is usually used for description of the MBA's size-distribution. It can obtain from (2) and (3) the next expression

$$\lg N = a_1 H + a_0 \qquad (4)$$

$N$ – quantity of all existing Centaurs with absolute magnitude less than $H$, $a_1$, $a_0$ – constant parameters.

Dependence $lgN(H)$ for all Centaurs from the MPC catalogue is presented in fig.1 (points). The dotted line marks the corresponding theoretical dependence (4) (for all existing bodies). One can see from fig.1 that almost all Centaurs with $H < 13.25^m$ ($D > 10$ km) are already revealed. Range of $H > 13.25^m$ is the range with unrevealed bodies. Existence of the quasi-linear site ($H < 13.25^m$) in the dependences $lgN(H)$ for the catalogue orbits is a some confirmations of validity of formulae (3) using for Centaurs.

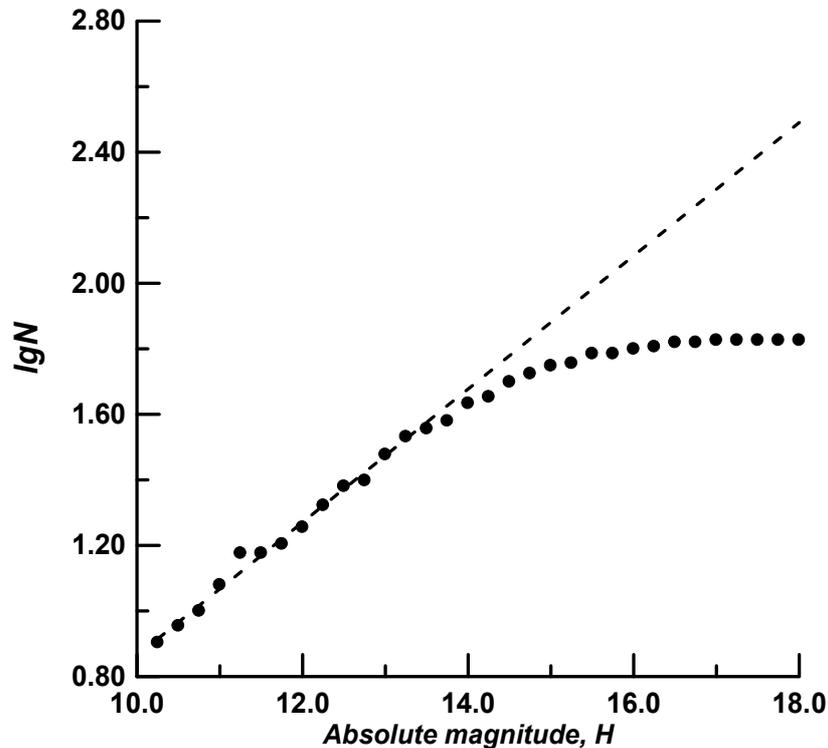

Fig. 1. Dependences $lgN(H)$ for Centaurs: points – for MPC catalogue orbits, dotted line - the theoretical dependence (for all existing bodies).

The total quantity of Centaurs with $H < 18.3^m$ ($D > 1$ km) ought be about 350. There were added 280 artificial Centaurs to the real bodies. Four model orbits corresponded to one of real Centaur's orbit. All orbital elements of artificial Centaurs corresponded to the real ones, with the exception of the mean anomaly. Mean anomalies of all artificial orbits had four values only: $0°$, $90°$, $180°$, and $270°$.

Thus we obtain the set of JFCs with 210 orbits, the set of MdPCs with 422 orbits and the set of Centaurs with 350 orbits.

## 4. Orbital evolution. The integration method and the results

The calculations were carried out by numerical integration of equations of motion in rectangular coordinates by on the method described in the paper (Kazantsev 2002). There were taken into account the influence of 8 planets, Pluto, Ceres and two the largest asteroids (2 and 4). Relativistic effects in perihelion motions were taken into account as well.

Integration intervals for all bodies amounted to 100000 years. Non-gravitational forces weren't taken into account. It is impossible to know course of non-gravitational forces for such long periods. But it doesn't mean that the main conclusions, based on the calculation results are wrong. No doubt, we can't calculate exactly orbit evolution for a separate comet for a long time interval without taking non-gravitational forces into consideration. Nevertheless, common evolution regularities for small body orbits can be revealed from integration of many orbits. After all, the dominant force in orbit evolution of Solar system small bodies (sizes $D > 1$ km) is the gravitation.

During the calculation there were registered all bodies which leaved from JFCs and which passed into the family. If aphelion distance $Q$ of a separate orbit of JFCs becomes great than 6.1 AU, the comet was regarded as a body out of the family. If aphelion distance $Q$ of a separate orbit of MdPCs or of Centaurs becomes less than 6.0 AU, the body was regarded as a member of the JFCs. Besides all movings to long-period orbits and to parabolic or hyperbolic orbits were registered as well.

The quantities of these groups at different moments are pointed in Tabl.2. The table includes such data: $T$ – time interval since the integration beginning, $Ncjr$ – comet quantity which remained in the Jupiter's family, $Ncjo$ – comet quantity which left the family, $Nmp$ –comet quantity which replenished JFC from MdPCs, $Nct$ –quantity of bodies which replenished JFC from Centaurs, $Nmpct$ – total quantity of bodies which replenished JFC from MdPCs and Centaurs.

Table 2. Data for small body quantities of different groups at different integration periods

| T,yeas | Ncjr | Ncjo | Nmp | Nct | Nmpct |
|---|---|---|---|---|---|
| 0 | 210 | 0 | 0 | 0 | 0 |
| 20 | 205 | 5 | 3 | 0 | 3 |
| 40 | 204 | 6 | 4 | 0 | 4 |
| 100 | 188 | 22 | 6 | 0 | 6 |
| 200 | 175 | 35 | 8 | 0 | 8 |
| 500 | 152 | 58 | 17 | 0 | 17 |
| 1000 | 132 | 78 | 22 | 0 | 22 |
| 2000 | 111 | 99 | 35 | 3 | 38 |
| 5000 | 87 | 123 | 37 | 5 | 42 |
| 10000 | 75 | 135 | 32 | 8 | 40 |
| 20000 | 44 | 166 | 27 | 8 | 35 |
| 40000 | 35 | 175 | 23 | 9 | 32 |
| 60000 | 30 | 180 | 16 | 18 | 34 |
| 80000 | 26 | 184 | 11 | 4 | 15 |
| 100000 | 29 | 181 | 9 | 7 | 16 |

The data of Tabl. 1 are presented in Fig. 2. One can see from Tabl. 2 and Fig. 2., that replenishment rate of JFCs from the outer sources is 3-5 times less than rate of comet outflow from the family at all stages of integration. Replenishment rate of JFCs from the model MdPCs is approximately the same as the replenishment rate from the real comets. It means the adding of the model orbits makes the quantitative changes only in the MdPC population but not the qualitative ones. On the base of the above data (Tabl. 2 and Fig. 2) one may draw a conclusion that replenishment of JFCs should mainly occur from the interior sources.

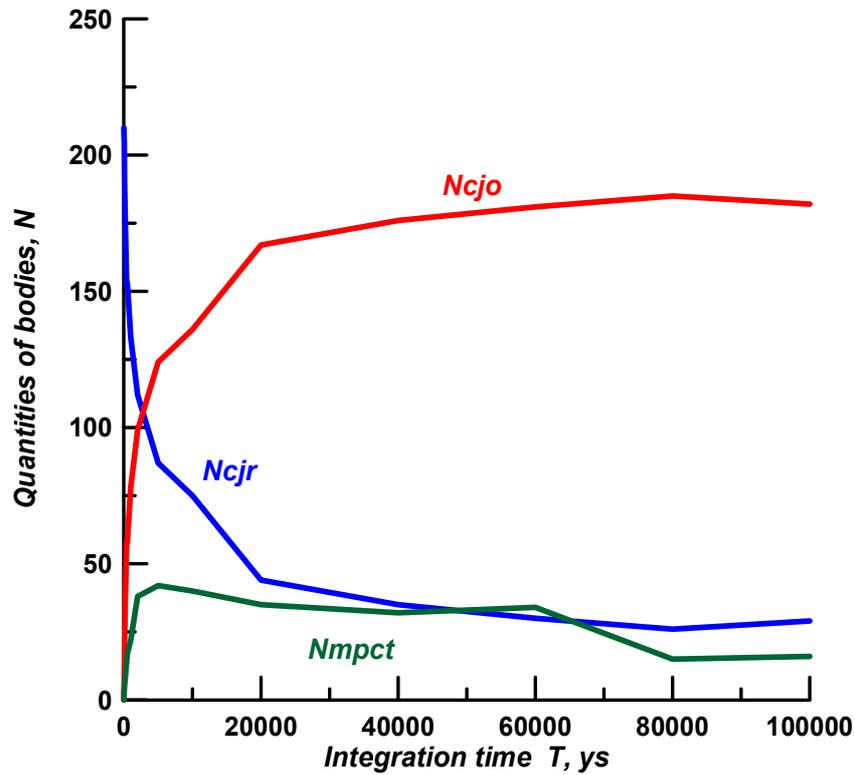

Fig.2. Outflow and replenishment of JFCs.

It is significant, that all comets from JFC which remained in the family are moving or in one of commensurabilities with Jupiter, either on orbits with aphelion distances Q < 4.6 AU. Consequently such comets can keep one's distance from Jupiter. If non-gravitational forces are operating, the comets can leave commensurabilities and approach to Jupiter.

The above result about more intensive outflow of JFCs in compare with their replenishment from the outer sources does not point to a special situation of Jupiter in the space. The dominant increase of semi-major axes is typical for small bodies moving beyond Jupiter's orbit as well. Data on changes of semi-major axes for MdPCs and Centaurs are pointed in Tabl. 3. The table includes such data: $T$ – time interval since the integration beginning, $Na+$ – the percent orbit quantity with increase of semi-major axis relatively the initial value, $Na-$ – orbit quantity with decrease of $a$, $Np$ – orbit quantity which moved to the parabola or the hyperbola.

Table. 3. The quantities of MdPCs and Centaurs with semi-major exes increase and decrease

| T, yeas | MdPCs | | | Centaurs | | |
|---|---|---|---|---|---|---|
| | $Na+$ | $Na-$ | $Np$ | $Na+$ | $Na-$ | $Np$ |
| 400 | 52.8 | 47.2 | 0 | 50.3 | 49.7 | 0 |
| 10000 | 64.0 | 36.0 | 23 | 54.0 | 46.0 | 12 |
| 100000 | 84.4 | 15.6 | 152 | 69.7 | 30.3 | 87 |

One can see from the table the $Na+$ are increasing in time. It means that the dominant increase of semi-major axes of small bodies, which can approach to the giant planets, is a nature phenomenon.

If the numbering estimation of all existing Centaurs have a considerable error, it haven't an influence on the validity of the results. First, as one can see from tabl. 2, Centaurs make a little contribution to the population of JFCs. It can explain by specific values of perihelion distances of Centaur's orbits (Kazantsev 2010).

Second, a little part of Centaurs which move to the orbits of JFCs, become the comets. About 60% of Centaurs in the catalogue have the orbit perihelion distances $q < 3$ AU, but don't show any comet activity. Consequently the main outer source of JFCs is population of MdPCs. Inflow of those bodies into Jupiter's family is 4-5 times less than comet outflow from the family.

## 5. A qualitative interpretation of the above results

The conclusion of dominant increase of semi-major axes of small bodies can be explained by the mechanism of planet action on a small body. Under an approach a planet, the planet-centric velocity vector of the small body turns without changing of module of the velocity vector. At that module of the heliocentric velocity vector $\mathbf{v}_h$ of the small body changes by value $\Delta v_h$. Probability and extent of the module increase is approximately the same as probability and extent of the module decrease.

Heliocentric distance $r$ and heliocentric velocity of a small body $\mathbf{v}_h$ are connected with the semi-major axis $a$ by the integral of energy

$$\mathbf{v}_h^2 = 2/r - 1/a \qquad (5)$$

At the increasing of modulus of $\mathbf{v}_h$ the semi-major axis $a$ is increasing, and vice versa. If modulus of the velocity changes in a range of $\pm \Delta v_h$, the extent of increase of the semi-major axis is greater than the extent of the decrease. The difference between the increase and the decrease of $a$ is expanding at the extension of range of $\Delta v_h$. A respective dependences $a(\Delta v_h)$, obtained from equation (5), are shown in Fig. 3. The upper line corresponds to the positive values of $\Delta v_h$, the lower line corresponds to the negative ones. For example a typical orbit of JFC with elements: $a = 6.62$ AU, $e = 0.44$ was chosen here. The model approach Jupiter occurs at heliocentric distance of 5.2 AU.

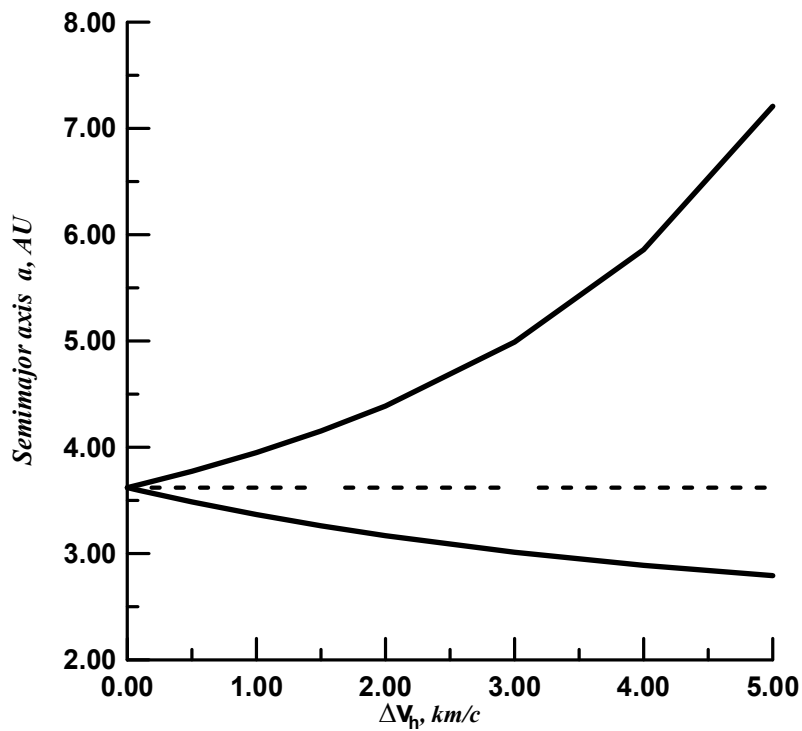

Fig.3. Changing of semi-major axis of small body depending on increase and decrease of heliocentric velocity under approaches the planet (Jupiter)

Consequently the numerical calculation results agree with the mechanism of planet action to a small body under the approaches.

# 6. Conclusions

1. Orbital evolution of the solar system small bodies is mainly going in direction of the semi-major exes increase. It belongs to the bodies which can undergo approaches the planets, and orbital evolution of which is mainly going due to the gravitational forces.

2. The main source of nuclei of Jupiter's family comets is apparently situated at distances from the Sun not more than 6 AU.

It is likely the conclusions seem wrong with account of the dominant opinion on the JFCs origin. In that case it is more important to try to find errors in the calculations and in interpretations, than to disregard ones.